\documentstyle[aps,prl,twocolumn,epsfig]{revtex}
\draft

\newcommand{\ba}{\begin{eqnarray}}
\newcommand{\ea}{\end{eqnarray}}
\def\ii{\'{\i}}
\begin{document}
\pagestyle{plain}

\title{Comment on `Two-body random ensembles: 
from nuclear spectra to random polynomials'}
\author{R. Bijker$^{a}$ and A. Frank$^{a,b}$}
\address{$^{a}$Instituto de Ciencias Nucleares, 
Universidad Nacional Aut\'onoma de M\'exico, \\
Apartado Postal 70-543, 04510 M\'exico, D.F., M\'exico
\newline
$^{b}$Centro de Ciencias F\ii sicas, 
Universidad Nacional Aut\'onoma de M\'exico, \\
Apartado Postal 139-B, Cuernavaca, Morelos, M\'exico}

\maketitle


An investigation of the spectroscopy of even-even nuclei with 
random one- and two-body interactions has led to the surprising 
result that the ground state has $J^P=0^+$ in typically $60-70$ 
$\%$ of the cases. This holds both for the nuclear shell model 
\cite{JBD,BFP} and the interacting boson model \cite{BF1}. 
To understand this phenomenon it was suggested to study the 
distribution of the lowest eigenvalues \cite{BF2}. To this end, 
a technique was developed in \cite{DK} for tridiagonal 
Hamiltonian matrices, such as for the one- and two-body 
interactions in the vibron model. It relies on the property 
that for these matrices it is possible to obtain an estimate for 
the lowest eigenvalue as the minimum of a function \cite{Hollenberg}. 

The purpose of this comment is to point out that for 
the vibron model the appropriate function is not a 
quadratic polynomial as used in \cite{DK}, but a more 
general form which in the vibrational basis to leading order 
in $1/N$, is given by 
\ba
f(z,j) = a \, z^2 + b \, z + c + d \, j^2 
+ e \, \sqrt{(1-z)^2(z^2-j^2)} ~. 
\nonumber
\ea
Here $z=n_p/N$ and $j=J/N$ \cite{DK}. 
In the absence of off-diagonal terms ($e=0$), the spectrum 
is that of a three-dimensional oscillator characterized by 
$n_p=0,1,\ldots,N$ and $J=n_p,n_p-2,\ldots,1$ ($n_p$ odd) 
or $0$ ($n_p$ even). In this case, the function $f(z,j)$ reduces 
to a quadratic polynomial, and the results of \cite{DK} can be used 
to obtain the percentages for ground states with $J=0$, $J=1$ and 
$J=N$ to be $64.6$, $5.2$ and $19.8$ $\%$, respectively, in 
close agreement with the exact values of $64.4$, $4.1$ and 
$24.2$ $\%$ obtained for $N=20$ and 100000 runs. 
In the case of only off-diagonal interactions ($a=b=c=d=0$), 
the ground state always has $J=0$. 

In the general case, both diagonal and off-diagonal terms 
are present. The off-diagonal term introduces a deformation 
in the system, which gives rise to the occurrence of vibrational 
bands with $J=0,1,2,\ldots,J_{\rm max}$. As a consequence, the 
$J=1$ ground states tend to disappear in favor of an enhancement of  
the $J=0$ ground states. This effect explains qualitatively 
the change in the $J=0$, $J=1$ and $J=N$ ground states from 
$64.4$, $4.1$ and $24.2$ to $72.1$, $0.5$ and $23.7$ $\%$, 
respectively. 

In order to carry out a more quantitative analysis, we 
have developed a simple method to analyze arbitrary 
functions $f(z,j)$ by sampling the two-dimensional space 
($z,j$). As an example, in Figure~1 we show the percentages of 
ground states with $J=0$ as a function of 
$x=\sigma_1/(\sigma_0+\sigma_1)$, where $\sigma_i$ denotes 
the width of the Gaussian distributions for the diagonal 
($i=0$) and off-diagonal ($i=1$) matrix elements. 
The case of equal 
widths (discussed in the previous paragraph) corresponds 
to $x=0.5$. Figure~1 shows a good agreement between the 
values obtained from an analysis of $f(z,j)$ (dashed line) 
and the exact results (solid line). The discrepancies observed 
for the quadratic polynomials of \cite{DK} (dotted line)
are due to a different treatment of the off-diagonal matrix element. 

In conclusion, the approach proposed in \cite{DK} provides an 
interesting method to analyze random interactions in terms of 
random polynomials which is especially suited for tridiagonal 
Hamiltonian matrices. The advantage of this method is that it 
foregoes the diagonalization of thousands of matrices. 
In this comment we have presented a careful analysis of the 
dominance of $J^P=0^+$ ground states in the vibron model, and 
found good agreement with the exact calculations. 

For cases of interest to nuclear physics, such as the IBM or 
the shell model, the Hamiltonian matrix has a more complicated form. 
Although it can be brought to tridiagonal form, the lack of an 
analytic form for the function $f$ leads to complications comparable 
to those of bringing it to diagonal form right away. For these cases 
the method becomes difficult to apply (other than by analogy). 
Many questions remain open to explain the robust features 
observed in randomly interacting many-body systems. 

This work was supported in part by CONACyT under projects 
32416-E and 32397-E, and by DPAGA under project IN106400.

\begin{figure}[h]
\centerline{\hbox{
\epsfig{figure=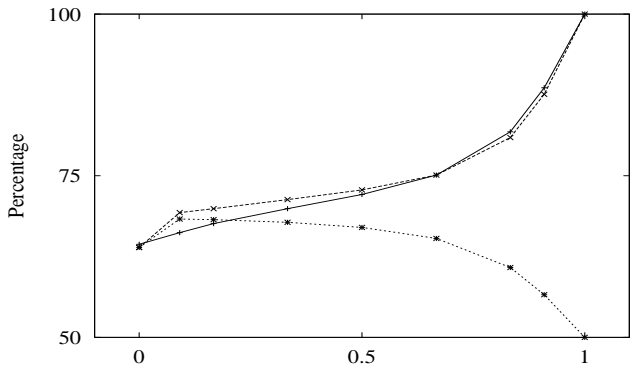,height=0.3\textwidth,width=0.5\textwidth} }}
\caption[]{Percentage of $J^P=0^+$ ground states as a function 
of $x$ (defined in text) for the exact results (solid), for 
$f(z,j)$ (dashed) and for the quadratic polynomial (dotted) 
\protect\cite{DK}.}
\end{figure}

\end{document}